\documentclass[amsmath,amssymb]{revtex4}

\usepackage{graphicx}% Include figure files
\usepackage{dcolumn}% Align table columns on decimal point
\usepackage{bm}% bold math

\begin{document}
\title{Gamma-ray emission in alpha-particle reactions with C, Mg, Si, Fe}

\author{J. Kiener$^1$, J. Bundesmann$^2$, I. Deloncle$^1$, A. Denker$^2$,  V. Tatischeff$^1$, 
A. Gostojic$^1$,  C. Hamadache$^1$, J. R\"{o}hrich$^2$, H. Benhabiles$^3$, 
I. Bourgaoub$^1$, A. Coc$^1$, F. Hammache$^4$, R. Mezhoud$^3$, J.Peyr\'e$^1$}
\address{$^1$CSNSM Orsay, CNRS/IN2P3 and Univ. Paris-Sud, France}
\address{$^2$Protonen in de Therapie, Helmholtz-Zentrum Berlin, Germany}
\address{$^3$Univ. M'HAMMED BOUGARA Boumerd\`es, Algeria}
\address{$^4$IPN Orsay, CNRS/IN2P3 and Univ. Paris-Sud, France}

\email{jurgen.kiener@csnsm.in2p3.fr}

\date{\today}

\begin{abstract}
Cross sections  for the strongest $\gamma$-ray emission lines produced in $\alpha$-particle reactions with C, Mg, Si, Fe have been measured in the range $E_{\alpha}$ = 50 - 90 MeV at the  center for proton therapy at the Helmholtz-Zentrum Berlin. Data for more than 60 different $\gamma$-ray lines were determined, with particular efforts for lines that are in cross section compilations/evaluations with astrophysical purpose, and where data exist at lower projectile energies. The data are compared with predictions of a modern nuclear reaction code and cross-section curves of the latest evaluation for gamma-ray line emission in accelerated-particle interactions in solar flares. 
\end{abstract}

\maketitle

\section{Introduction}
Cross sections for the emission of nuclear $\gamma$-ray lines are a basic ingrediant for the analysis and interpretation of observations in the low-energy $\gamma$-ray band of astrophysical sites with an important population of accelerated ions.  The latest compilation of Murphy et al.  \cite{MKKS}, aimed mainly at de-excitation lines produced in accelerated-particle interactions in solar flares, features 248 cross-section curves for $\gamma$-ray lines emitted in proton, $^3$He and $\alpha$-particle reactions  with abundant isotopes in the solar atmosphere.  The curves are typically given from reaction threshold to several hundred MeV per nucleon, which is the important energy range for the accelerated particle populations in solar flares. For most of the curves, however, data exist only at proton energies below about 25 MeV and for $\alpha$-particles below about 10 MeV per nucleon and the extrapolation to higher energies relied on calculations with the nuclear reaction code TALYS \cite{Talys}. 

Of particular importance are the most prominent lines that often stand out in observed spectra from strong solar flares. They can be used to deduce ambient abundances and the energetic-particle composition and energy spectrum \cite{Osse_flare, MRKR}.  These lines are  from the deexcitation of the first or second excited state in the most abundant species: $^{12}$C, $^{14}$N, $^{16}$O, $^{20}$Ne, $^{24}$Mg, $^{28}$Si, $^{56}$Fe. The cross section curves for the emission of these lines in proton and $\alpha$-particle interactions with these nuclei typically show a broad maximum around 15-20 MeV, due to inelastic scattering reactions, reaching the several hundred mb range. This is followed by an nearly exponential fall-off to higher energies for proton reactions while a second, very broad maximum around 60 MeV is predicted for $\alpha$-particle reactions. This maximum may be explained by fusion-evaporation reactions like ($\alpha$,dnp) or ($\alpha$,2n2p), where the residual nucleus is identical to the target nucleus.

This second cross section maximum is important in solar flares with relatively high accelerated $\alpha$-particle to proton ratios $\alpha$/p and hard energy spectra. However, there are practically no experimental data available for the emission of $\gamma$-ray lines in $\alpha$-particle reactions above $E_{\alpha}$ = 40 MeV. We therefore decided to measure cross sections for the emission of strong lines in $\alpha$-particle reactions around the predicted second cross section maximum. In the first experiment, we concentrated on the elements C, O, Mg and Fe. The chosen energy range, $E_{\alpha}$ = 50 - 90 MeV also continues to higher energies the data ($E_{\alpha}$$\leq$ 40 MeV), obtained in previous experiments at the Orsay tandem-van-de-Graaff accelerator \cite{t2002, Gamflare}.

\section{Experiment}

The experiments have been done at the center for proton therapy of the Helmholtz-Zentrum Berlin in two campaigns in 2015 and 2016. Pulsed beams of $\alpha$ particles with $E_{\alpha}$ = 50 MeV in 2015 and  $E_{\alpha}$ = 60, 75 and 90 MeV in 2016 have been produced using the Van-de-Graaff injector and the K-132 separated sector cyclotron of the Helmholtz center. They were directed onto self-supporting target foils of C, Mg, Si and Fe inside a spherical chamber of 15 cm diameter made of stainless steel and stopped in a thick copper block about 4 m downstream of the target chamber. The target chamber was equipped with two glass windows perpendicular to the beam direction, where one of them served for optical monitoring of the beam spot through the use of an alumina target. In both campaigns, 4 detectors, one LaBr3:Ce scintillation detector and three HP-Ge detectors in a wide angular range and at large distances of typically 50-70 cm from the target, have been used for the detection of the emitted $\gamma$ rays. 

The pulsed beams were produced with relatively low repetition rates of $\sim$80 kHz to $\sim$1 MHz and a small pulse width well below 1 ns. This allowed an efficient suppression of background, in particular radiation from the beam stop and interactions of secondary neutrons in the detectors, but also radiation from activated material and natural radioactivity. Beam intensities were of the order of 0.5-3 nA, resulting in typical detector count rates of a few thousand counts per second. The beam current was repeatedly measured in two well-shielded Faraday cups sitting just after the beam extraction from the cyclotron and about 1 m upstream of the targets. Values of the two Faraday cups agreed to better than 5\%. In 2016, a third cup was installed at the beam stop, allowing a constant monitoring of the beam intensity. The efficiency calibration of the detectors has been done with standard radioactive sources, $^{133}$Ba, $^{137}$Cs, $^{60}$Co and $^{152}$Eu for energies below about 1.4 MeV and extensive Geant simulations to extrapolate the detector efficiencies to higher energies and also to obtain precise detector response functions.

\begin{figure}[h]
\includegraphics[width=30pc]{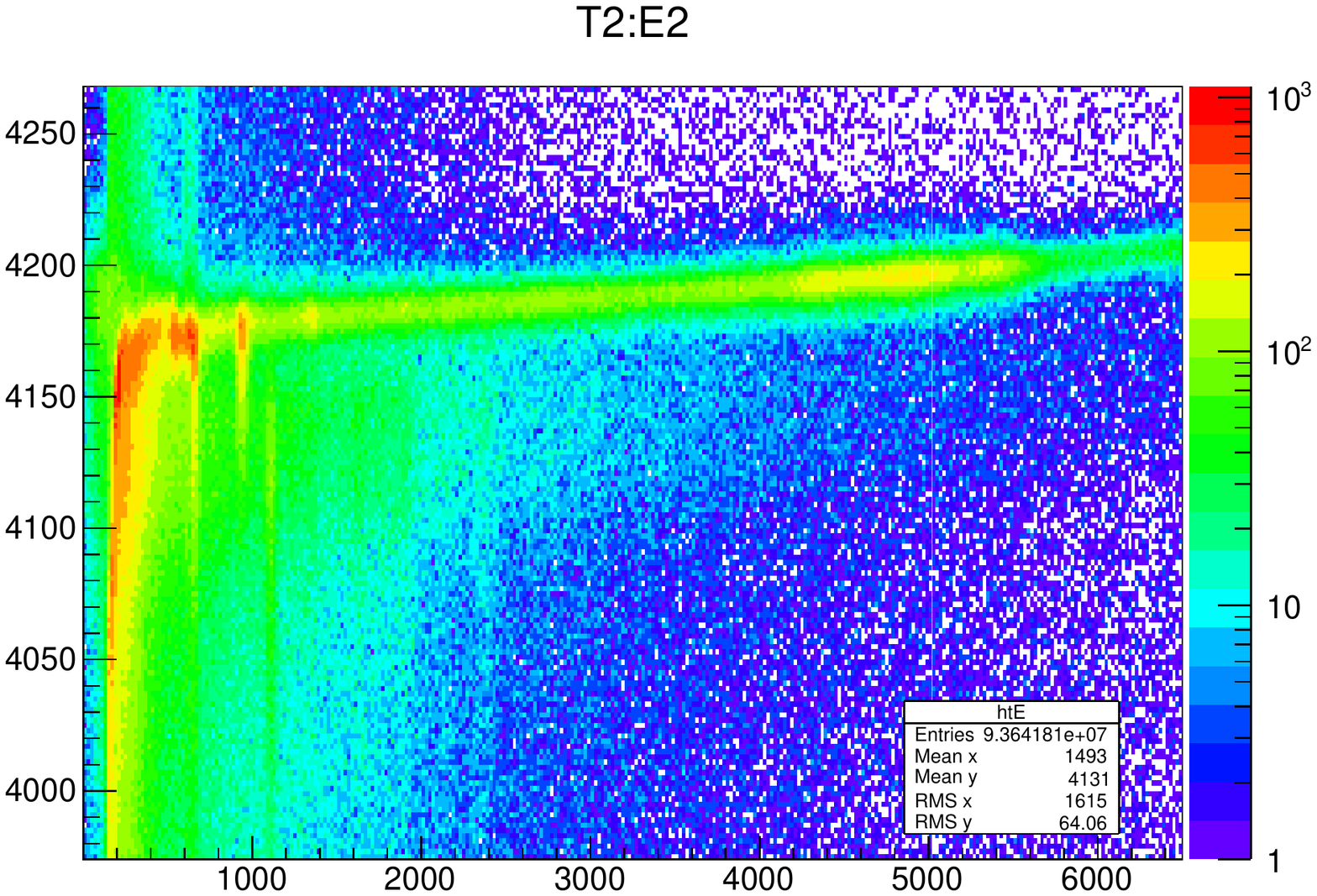}
\caption{\label{label}Timing channel (T2) versus energy channel (E2) of events in the LaBr$_3$:Ce detector recorded during the irradiation of the C target with a 90-MeV $\alpha$-particle beam.}
\end{figure}

The detector signals were treated with standard NIM modules, providing a signal for the deposited energy and another signal for the timing of the event. The timing signal was defined as the time difference between the detector signal and the beam pulse signal, both determined with constant fraction discriminators fed into a time-to-amplitude converter NIM module. The energy and timing data of the 4 detectors were then digitized and recorded in event-by-event mode and a time stamp of 100 ns resolution using an acquisition system of FAST ComTec \cite{FAST}, providing also the dead times of the 4 energy and 4 timing channels.

Figure 1 shows a zoom in the timing vs. energy plane of data recorded by the LaBr$_3$:Ce detector in the irradiation of the C target with a 90-MeV $\alpha$-particle beam. The slightly inclined, horizontal branch at T2 $\sim$4190 shows the prompt $\gamma$ rays induced by beam interactions in the target. At E2 $\sim$4200-5400 is the 4.4-MeV complex including the full-energy and escape lines, some narrow lines can also be seen below E2 $\sim$1500. A broader horizontal band inside T2 $\sim$4100-4160 with several narrow lines below E2 $\sim$2400 can also be recognized. These events, arriving about 4 ns after the prompt target $\gamma$ rays, have been identified to come from interactions of secondary particles, mostly neutrons probably, inside the stainless-steel walls of the reaction chamber. Events arriving still later are from interactions in further away beam tubes and supporting structures and from secondary neutrons interacting in the LaBr$_3$:Ce crystal. Beam-induced $\gamma$ rays from interactions in the beam stop are in this histogram around T2 $\sim$2700 (not shown).

\begin{figure}[h]
\includegraphics[width=30pc]{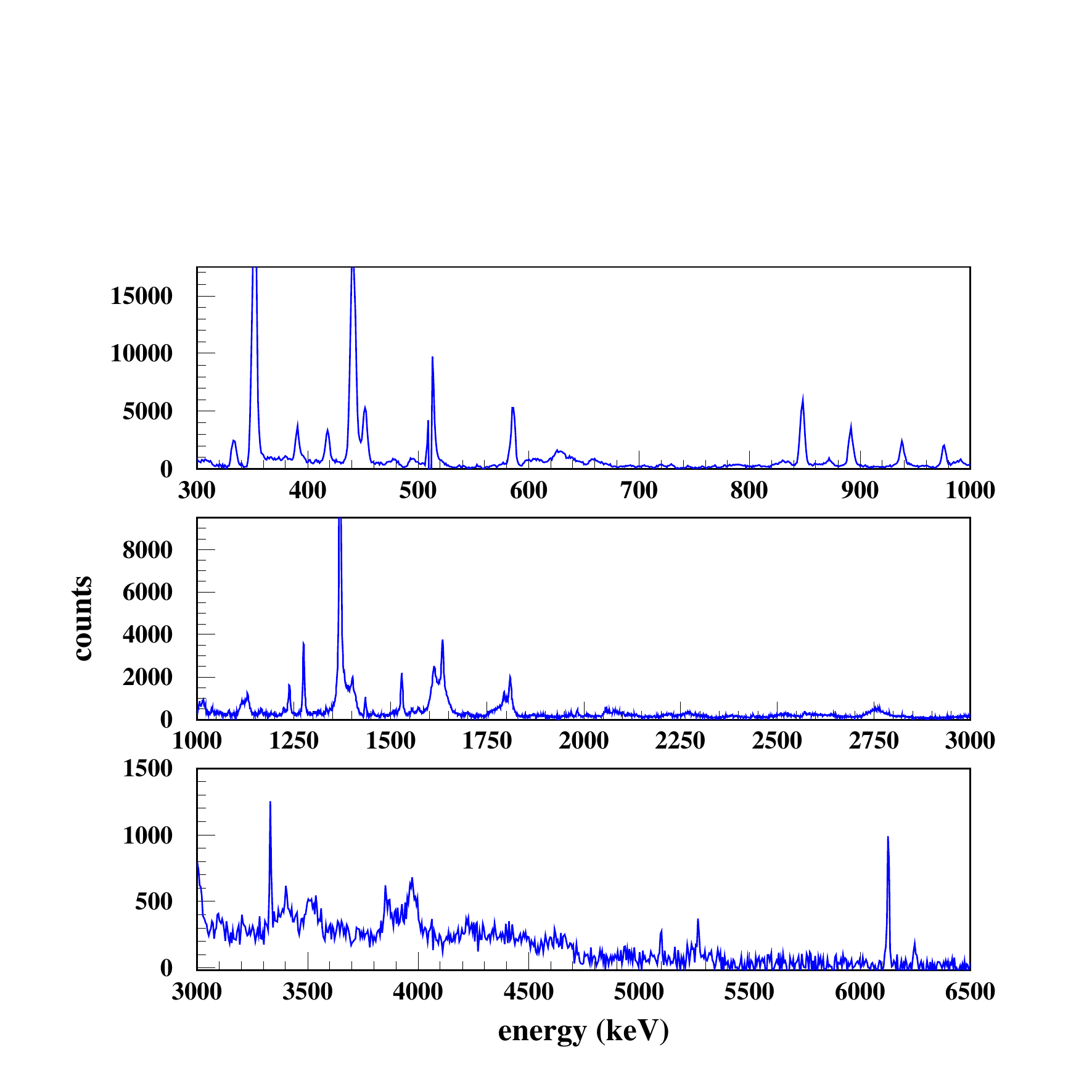}
\caption{\label{label}Compton-subtracted spectrum of the HP-Ge detector at 90$^o$ to the beam direction from irradiation of the Mg target with 75-MeV $\alpha$ particles.}
\end{figure}

\section{Data analysis and results}

Spectra of prompt $\gamma$ rays from the target were obtained by a selection of events within a narrow band around the visible horizontal branchs in the energy-timing plane from prompt target $\gamma$ rays as shown in Figure 1. With the LaBr$_3$:Ce detectors, events from interactions of secondary particles could be completely excluded by the selection, while for the HP-Ge detectors, with a time resolution not better than 3-4 ns, not all events from secondary-particle interactions in the target chamber walls could be separated. Their contributions, essentially lines from the first few excited states of $^{56}$Fe and $^{52}$Cr, could however be estimated from their time profile in the LaBr$_3$:Ce detector. This was important for cross-section determinations with the Fe target. 

Integrals for narrow $\gamma$-ray lines could be directly taken from these spectra, while for broader lines and line complexes, Compton-subtracted spectra were used. These spectra were obtained by an unfolding procedure made possible by dedicated measurements with radioactive sources and extensive Geant simulations for the detector response functions in the two experimental setups. The accuracy of the Compton subtraction from about 8 MeV down to the 511-keV line was estimated to be better than 20\%. An example of Compton-subtracted spectrum is shown in Figure 2. 

Some of the most important lines for each target, broad lines (like $E_{\gamma}$ = 4439 keV in $^{12}$C) or narrow ones (e.g. $E_{\gamma}$ = 1368 and 1779 keV in $^{24}$Mg and $^{28}$Si, respectively) were merged with other lines in a complex structure and needed line-shape calculations to determine their intensity. Examples for the 1779-keV line and the 4439-keV line are shown in Figures 3, 4. The uncertainties related to these decompositions is added to the statistical and other systematic uncertainties like the detector efficiencies. In cases of lines very close in energy, like the 4439-keV line of $^{12}$C  and the 4445-keV line of $^{11}$B, the sum has been used for the line cross section determination.

\begin{figure}[h]
\begin{minipage}{16pc}
\includegraphics[width=16pc]{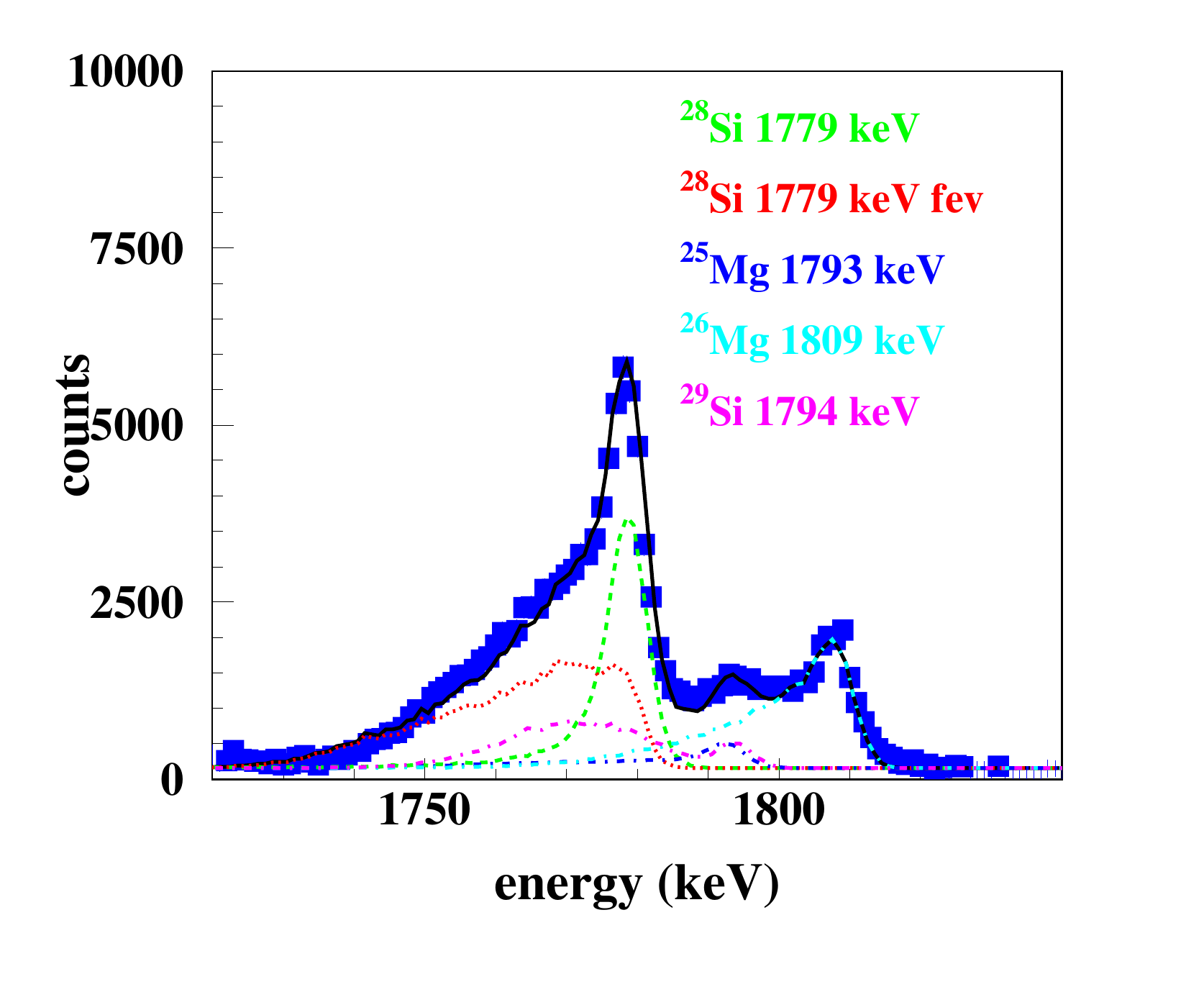}
\caption{\label{label}Symbols show the line complex with the 1779-keV line of $^{28}$Si in the Compton-subtracted spectrum of the HP-Ge detector situated at 147$^o$ to the beam direction from irradiation of the Si target with 60-MeV $\alpha$ particles. The line "$^{28}$Si 1779 keV fev" on the figure means production of the 1779-keV line by fusion-evaporation reactions. }
\end{minipage}\hspace{2pc}%
\begin{minipage}{16pc}
\includegraphics[width=16pc]{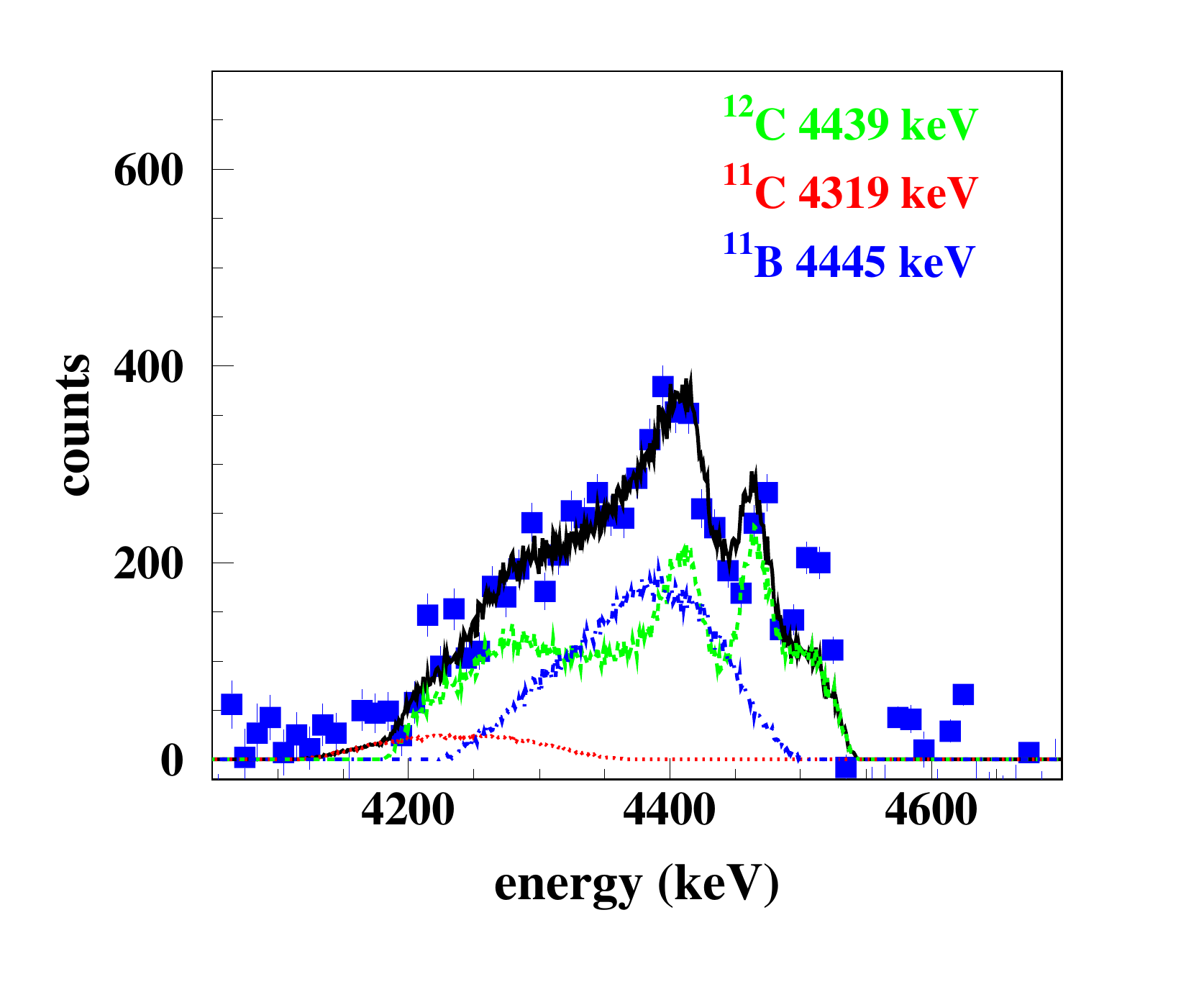}
\caption{\label{label}Symbols: Compton-subtracted spectrum around the 4.4-MeV line complex of the HP-Ge detector at 116$^o$ to the beam direction during irradiation of the C target with 50-MeV $\alpha$ particles. The shape of the 4439-keV line of $^{12}$C has been obtained using coupled-channels calculations for inelastic scattering as in \cite{Kien19}.  }
\end{minipage} 
\end{figure}

Differential cross section data from the 3 HP-Ge detectors have been obtained for about 60 different $\gamma$-ray lines, and for most of them at all $\alpha$-particle energies. With the LaBr$_3$:Ce detectors, however, owing to their lower energy resolution, no cross section data could be obtained for some lines in complex structures.  Finally, the $\gamma$-ray emission cross sections have been obtained by Legendre-polynomial fits of the 3 or 4 detector data. To the resulting cross-section uncertainty from the fit were added the target thickness and beam charge uncertainties.

Cross section curves for the strongest line of each target, here from the deexcitation of the first excited level in the  major isotopes $^{12}$C, $^{24}$Mg, $^{28}$Si and $^{56}$Fe are shown on Figure 5. The data show effectively clearly a second cross section bump for the 1.78-MeV $^{28}$Si and 0.85-MeV $^{56}$Fe lines, approximately at the energies predicted by the compilation \cite{MKKS}.  The latter, adjusted to the data at lower energies, that were obtained at the tandem-Van-de-Graaff accelerators of Washington \cite{Dyer85,Seamster84} and Orsay \cite{t2002}, however, overestimates significantly the measured cross sections. For the 1.37-MeV line of  $^{24}$Mg, there is a hint for a weak second cross section bump, in any case significantly less pronounced than predicted by the compilation. Its prediction for the cross section curve of the 4.44-MeV line of $^{12}$C agrees reasonably with the present data.

\begin{figure}
\begin{center}
\includegraphics[width=25pc]{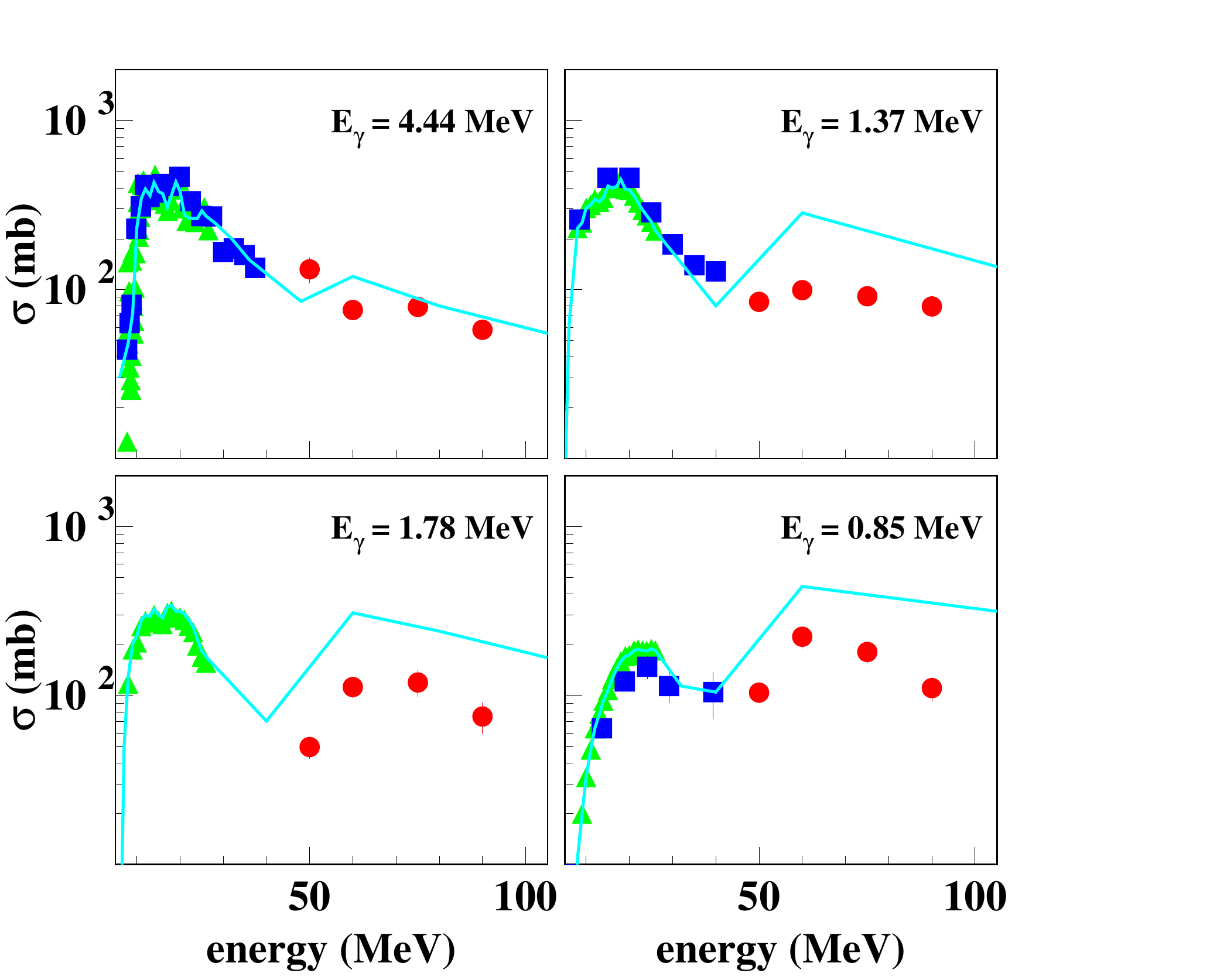}
\end{center}
\caption{\label{label}Gamma-ray line emission cross sections for $\alpha$-particle reactions with, from left to right and from upper to lower: C, Mg, Si and Fe. Red filled circles are the present data, blue filled squares are from Ref. \cite{t2002}, green triangles are from Ref. \cite{Dyer85} for C and from Ref. \cite{Seamster84} for Mg, Si and Fe.  The cyan curves are the cross section excitation functions from the cross section compilation \cite{MKKS}.}
\end{figure}

Similar disagreements can also be seen for cross sections of the other strong lines, albeit with a slightly better agreement for the lines from the Fe target compared to the other targets. We also did calculations with the nuclear reaction code Talys \cite{Talys} with default parameters for the nuclear structure and potentials. For the lines from the deexcitation of the first few levels of the major target isotopes, they generally agree at energies  below $E_{\alpha}$ $\sim$30 MeV with the data at the 30\% level or better, but disagree with the present data up to factors of 2-3. Here, a comprehensive study of structure parameters and nuclear reaction potentials in Talys as it has been done in \cite{t2002,Gamflare} could certainly improve significantly the agreement with data and allow an accurate description of the total $\gamma$-ray emission in $\alpha$-particle reactions and an extrapolation to higher projectile energies.

\section{Conclusions}

Cross section data for the emission of about 60 $\gamma$-ray lines could be determined for $\alpha$-particle reactions with targets of C, Mg, Si and Fe at $E_{\alpha}$ = 50, 60, 75 and 90 MeV. For the strongest lines from each target, there is now with the present data a complete coverage of experimental data from threshold to 90 MeV. Furthermore, cross sections for many new lines could be added, where no data were available from previous experiments. Comparison of cross section curves of the latest cross section compilation and nuclear reaction code calculations show significant disagreements with the new data. This underlines the importance and necessity of experimental work at particle accelerators for the establishment of an accurate cross section data base in a wide projectile energy range that is needed for applications in astrophysics, but also for other applications with energetic particles like hadrontherapy. New accelerator centers open possibilities for future measurements in astrophysics and nuclear physices \cite{TGG}.

\section{Acknowledgments}
We like to thank the staff at the section for proton therapy of the Helmholtz-Zentrum Berlin for their strong involvement to provide the high-quality beams in the time periods between the cancer therapy sessions and their generous and efficient help in the planning and setup of the experiments. 

\section*{References}
\bibliography{mybib}

\end{document}